\RequirePackage{fix-cm}
\documentclass[twocolumn,epjc3]{svjour3}  
\smartqed  
\RequirePackage{graphicx}
\usepackage{amssymb}
\usepackage{amsmath}

\newcommand{\tr}{{\rm tr} \,}

\newcommand{\Dslash}{D \hspace{-2.7mm}/ \;}

\journalname{Eur. Phys. J. C}

\begin{document}
\title{On a first order transition in QCD \\ with up, down and strange quarks}
\author{
Xiao-Yu Guo\thanksref{addr1} 
\and Yonggoo Heo\thanksref{addr2}  
\and Matthias F.M. Lutz\thanksref{e1,addr1,addr3}
}

\thankstext{e1}{e-mail: m.lutz@gsi.de}

\institute{GSI Helmholtzzentrum f\"ur Schwerionenforschung GmbH,
Planck Str. 1, 64291 Darmstadt, Germany \label{addr1}
\and
 Suranaree University of Technology, Nakhon Ratchasima, 30000, Thailand \label{addr2}
\and 
Technische Universit\"at Darmstadt, D-64289 Darmstadt, Germany \label{addr3}
}

\date{Received: date / Accepted: date}

\maketitle

\begin{abstract}
 We consider the quark-mass dependence of the baryon octet and decuplet ground state masses. 
It is predicted that QCD dynamics implies a first order transition when increasing the 
strange quark mass from its chiral limit towards its physical value. 
Our claim relies on a global fit to the available QCD lattice data on such baryon masses. 
Quantitative results 
based on an application of the chiral SU(3) Lagrangian at N$^3$LO are discussed. We  predict an anomalous sector of QCD where stable baryonic matter would be composed of $\Lambda$ or $\bar \Lambda$ particles rather than nucleons and anti-nucleons
\end{abstract}

\section{Introduction}

A most fundamental question of modern theoretical physics is to explain how small-scale structures are formed in terms of  local quantum field theories. Nature shows a plethora of phenomena, like finite nuclei, conventional or exotic hadrons, which are thought to be a consequence of the strong interaction as 
encoded in quantum chromodynamics (QCD). The non-perturbative nature of the latter makes it quite difficult to arrive at a quantitative link from QCD to observable quantities as measured in the laboratory (see e.g.~\cite{Guo:2017jvc,Lutz:2015ejy,Ryan:2016lml,Mohler:2017ibi}). While QCD lattice simulations are encouragingly successful by now to post- and predict the masses of hadrons in their ground states, it is still a long thorny path towards realistic computations 
of the QCD excitation spectrum (see e.g.~\cite{Lin:2008pr,Edwards:2011jj,Ryan:2015gda,Ryan:2016lml,Mohler:2017ibi}). 

Of particular interest are phase transitions that may arise in QCD as a function of temperature 
and/or baryon density (see e.g.~\cite{Fodor:1994sj,Aoki:2006br,Borsanyi:2010cj,Bhattacharya:2014ara}). While QCD lattice simulations have seemingly converged conclusions 
on a smooth crossover behaviour of the chiral transition at finite temperature, $T$, similar studies 
at finite baryon chemical potential, $\mu$, are still not possible (see e.g.~\cite{Fodor:2001pe,Fodor:2004nz}). 
In this Letter we wish to point at  another sector of QCD, which is within reach of current lattice QCD simulation technology. While QCD is expected to show a rich phase structure on the external parameters $T$ and $\mu$, possible parametric phase transitions in the up, down and strange quark masses did not 
receive much attention so far. The possibility of 
a discontinuous quark-mass dependence in the baryon masses was discussed in \cite{Semke:2006hd}. 
The purpose of this Letter is to take up this issue, given the rather large data set generated on various QCD lattice ensembles over the last 15 years. We argue that by now it is possible to arrive at more definite conclusions. 

QCD lattice simulations with three light flavors at pion and kaon masses smaller than $600$ MeV are considered that are available publicly. That leaves data sets from  PACS-CS, LHPC,  HSC,  NPLQCD, QCDSF-UKQCD and ETMC \cite{PACS-CS2008,LHPC2008,HSC2008,NPLQCD:2011,Bietenholz:2011qq,Alexandrou:2013joa}. 
We are aware of the recent lattice ensembles of the CLS group with 2+1 flavors based on nonperturbatively improved Wilson fermions \cite{Bruno:2014jqa,Bali:2016umi,Bruno:2016plf}. Results for 
baryon masses are not available yet.

\section{An effective field theory approach}

We study QCD in terms of its chiral SU(3) Lagrangian. The strong interaction of hadrons can be described efficiently in terms of effective degrees of freedom at least at low enough energies \cite{Gasser:1982ap,Gasser:1984gg}. The most prominent effective fields interpolate the pseudo Goldstone bosons, the pion, kaon and eta mesons. 
This sector  \cite{Gasser:1984gg,Bijnens:2006zp} of the chiral Lagrangian is well established 
\begin{eqnarray}
&& {\mathcal L} =  - f^2\,{\tr}  \{ U_\mu \,U^\mu \}
  +  {\textstyle{1\over 2}}\,f^2\,{\tr}  \big\{\,\chi_+  \big\}  
\nonumber\\
&& \quad - \,8\,L_4\,{\tr}\{U_\mu U^\mu\}\,{\tr}\{\chi_+\}- 8\,L_5\,{\tr}\{U_\mu U^\mu \chi_+\} 
\nonumber\\
&& \quad +\, 4\,L_6\,{\tr}\{\chi_+\}\,{\tr}\{\chi_+\} 
+\, \,4\,L_7\,{\tr}\{\chi_-\}\,{\tr}\{\chi_-\} 
\label{def-L1}\\
&& \quad +\,2\,L_8\,{\tr}\{\chi_+ \chi_+ + \chi_- \chi_- \}\,,\qquad \quad\; u = e^{i\,\frac{\Phi}{2\,f}} \,,
\nonumber\\
&& U_\mu = {\textstyle{1\over 2}}\,u^\dagger \, \big(
\partial_\mu \,e^{i\,\frac{\Phi}{f}} \big)\, u^\dagger
\,,\quad \chi_\pm = {\textstyle{1\over 2}}\, \big( u \,\chi_0 \,u \pm u^\dagger \,\chi_0 \,u^\dagger \big) \,, \nonumber
\end{eqnarray}
where we display terms only that are relevant for our current work. While the matrix field $\Phi$ in (\ref{def-L1}) combines the pion, kaon and eta meson fields into a suitable 3$\times 3$ matrix, the symmetry breaking field $ \chi_0 \sim {\rm diag}(m_u, m_d, m_s)$ is proportional to the current quark masses of QCD. By means of (\ref{def-L1}) the quark-mass dependence of the pion, kaon and eta meson masses can be computed in QCD. At the one-loop level besides the quark masses the low-energy constants $f, L_n$ are involved only. Here we use resummed expressions where the tadpole terms involve the on-shell meson masses \cite{Lutz:2018cqo,Guo:2018zvl}.

Accurate results at the 10 MeV uncertainty level for the baryon masses can be obtained at next-to-next-to-next-to-leading order (N$^3$LO). This involves a large set of low-energy constants (LEC). Here sum rules for the LEC as derived from QCD with a large number of colors $N_c$ are instrumental and reduce the number of fit parameters significantly \cite{Scherer:2009bt,Lutz:2010se,Lutz:2018cqo}. It is important to consider the baryon octet and baryon decuplet fields on an equal footing. After all in the large-$N_c$ limit of QCD the two flavor multiplets turn degenerate. 
Here we recall a selection of terms which involve the baryon octet fields only. Analogous terms with the decuplet fields can be taken from \cite{Lutz:2010se,Semke:2011ez,Lutz:2018cqo}. The chiral Lagrangian is 

{
\allowdisplaybreaks
\begin{eqnarray}
&& \mathcal{L} =
\mathrm{tr}\, \big\{ \bar B\, (i\, \Dslash\, - M )\, B \big\} 
+  F\, \mathrm{tr}\, \big\{ \bar{B}\, \gamma^\mu \gamma_5\,\big[i\,U_\mu,\,B \big]\, \big\}\nonumber\\
&&\qquad + D\, \mathrm{tr}\,\big\{ \bar{B}\, \gamma^\mu \gamma_5\, \big\{i\,U_\mu,\,B\big\}\, \big\}\nonumber \\
&&  \qquad +\,2\, b_0 \,\mathrm{tr}\, \big\{\bar B \,B\big\} \mathrm{tr}\,\big\{\chi_+\big\} + 2 \,b_D\,\mathrm{tr}\,\big\{\bar{B}\,\big\{\chi_+,\,B\big\}\big\} \nonumber \\
&&  \qquad+ 2\, b_F\,\mathrm{tr}\,\big\{\bar{B}\,\big[\chi_+,\,B\big]\big\}
\nonumber \\
&&  \qquad- \,{\textstyle{1\over 2}}\,g_0^{(S)}\,\mathrm{tr} \,\big\{\bar{B}\,B \big\}\, \mathrm{tr}\,\big\{ U_\mu\, U^\mu \big\}
\nonumber \\
&&  \qquad - {\textstyle{1\over 2}}\,g_1^{(S)}\,\mathrm{tr} \,\big\{ \bar{B}\, U^\mu \big\}\, \mathrm{tr}\,\big\{U_\mu\, B \big\}
\nonumber \\
&&  \qquad
-{\textstyle{1\over 4}}\,g_D^{(S)}\,\mathrm{tr}\,\big\{\bar{B}\big\{\big\{U_\mu, U^\mu\big\}, B\big\}\big\}
\nonumber \\
&& \qquad -\,{\textstyle{1\over 4}}\,g_F^{(S)}\,\mathrm{tr}\,\big\{ \bar{B}\big[\big\{U_\mu, U^\mu\big\}, B\big]\big\}
\nonumber \\
&&  \qquad
  -{\textstyle{1\over 4}}\,g_0^{(V)}\,  \mathrm{tr}\,\big\{\bar{B}\, i\,\gamma^\mu\, D^\nu B\big\} \,
\mathrm{tr}\,\big\{ U_\nu\, U_\mu \big\}
\nonumber \\
&& \qquad -\, {\textstyle{1\over 8}}\,g_1^{(V)} \,\big( \mathrm{tr}\,\big\{\bar{B}\,U_\mu \big\} \,i\,\gamma^\mu \, \mathrm{tr}\,\big\{U_\nu\, D^\nu B\big\} 
\nonumber \\
&& \qquad\qquad+ \mathrm{tr}\,\big\{\bar{B}\,U_\nu \big\} \,i\,\gamma^\mu \, \mathrm{tr}\,\big\{U_\mu\, D^\nu B\big\} \big)
\nonumber \\
&& \qquad -\, {\textstyle{1\over 8}}\,g_D^{(V)}\, \mathrm{tr}\,\big\{\bar{B}\, i\,\gamma^\mu \big\{\big\{U_\mu,\, U_\nu\big\}, D^\nu B\big\}\big\} 
\nonumber \\
&& \qquad -  {\textstyle{1\over 8}}\,g_F^{(V)}\, \mathrm{tr}\,\big\{ \bar{B}\, i\,\gamma^\mu\, \big[\big\{U_\mu,\, U_\nu\big\},\, D^\nu B \big]\big\}  \,,
\nonumber\\ 
\nonumber\\ 
&&\Gamma_\mu ={\textstyle{1\over 2}}\,u^\dagger \,\partial_\mu  \,u
+{\textstyle{1\over 2}}\, u \,\partial_\mu  \,u^\dagger\,,  
\nonumber\\ 
&& D_\mu \, B \;\,= \partial_\mu B +  \Gamma_{\mu}\, B -
B\,\Gamma_{\mu} \,.
\label{def-L2}
\end{eqnarray}
}

We turn to the main target of our Letter, a possible discontinuous quark-mass dependence of the baryon masses \cite{Semke:2006hd}. A conventional $\chi$PT approach will always lead 
to a smooth behaviour \cite{Young:2009zb,Ren:2012aj}. However, any such approach is at odds with the QCD lattice data sets on the baryon masses \cite{LHPC2008,PACS-CS2008,MartinCamalich:2010fp,WalkerLoud:2011ab}.  In a series of works \cite{Semke2005,Semke:2011ez,Semke:2012gs,Lutz:2014oxa,Lutz:2018cqo} it was demonstrated 
that significant results are achieved once on-shell masses are used in the loop contributions to the hadron masses. Most recently it was shown that such an approach can be cast into a form that not only 
leads to renormalization scale invariant results but also implies convincing convergence properties 
up to strange quark masses that enclose the physical point \cite{Lutz:2018cqo}. Given such a scheme, a set of non-linear equations has to be solved in order to arrive at the quark-mass dependence of the 
baryon masses. In turn there is no longer any reason to expect a smooth behavior everywhere. 

\begin{figure}
\centering
\includegraphics[width=0.43\textwidth]{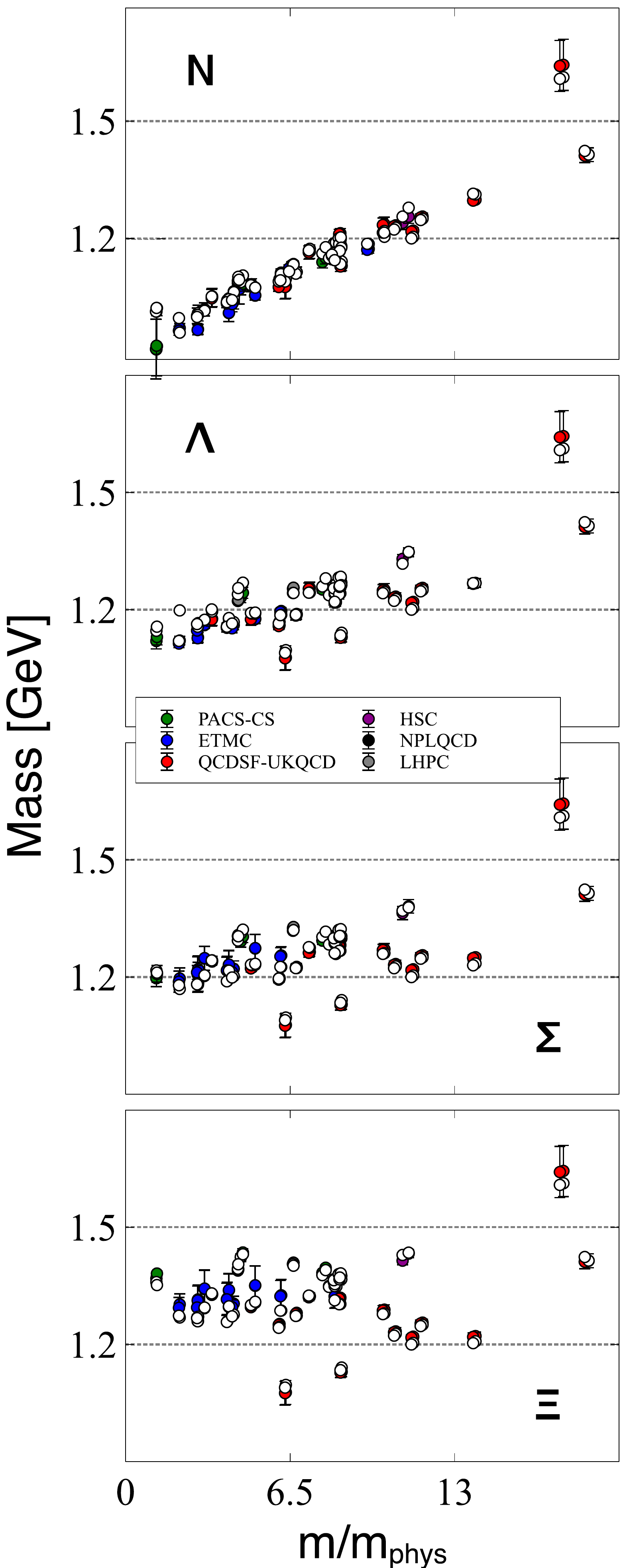}
\caption{Baryon octet masses as a function of the light quark mass as explained in the text.}
\end{figure}

\begin{figure}
\centering
\includegraphics[width=0.43\textwidth]{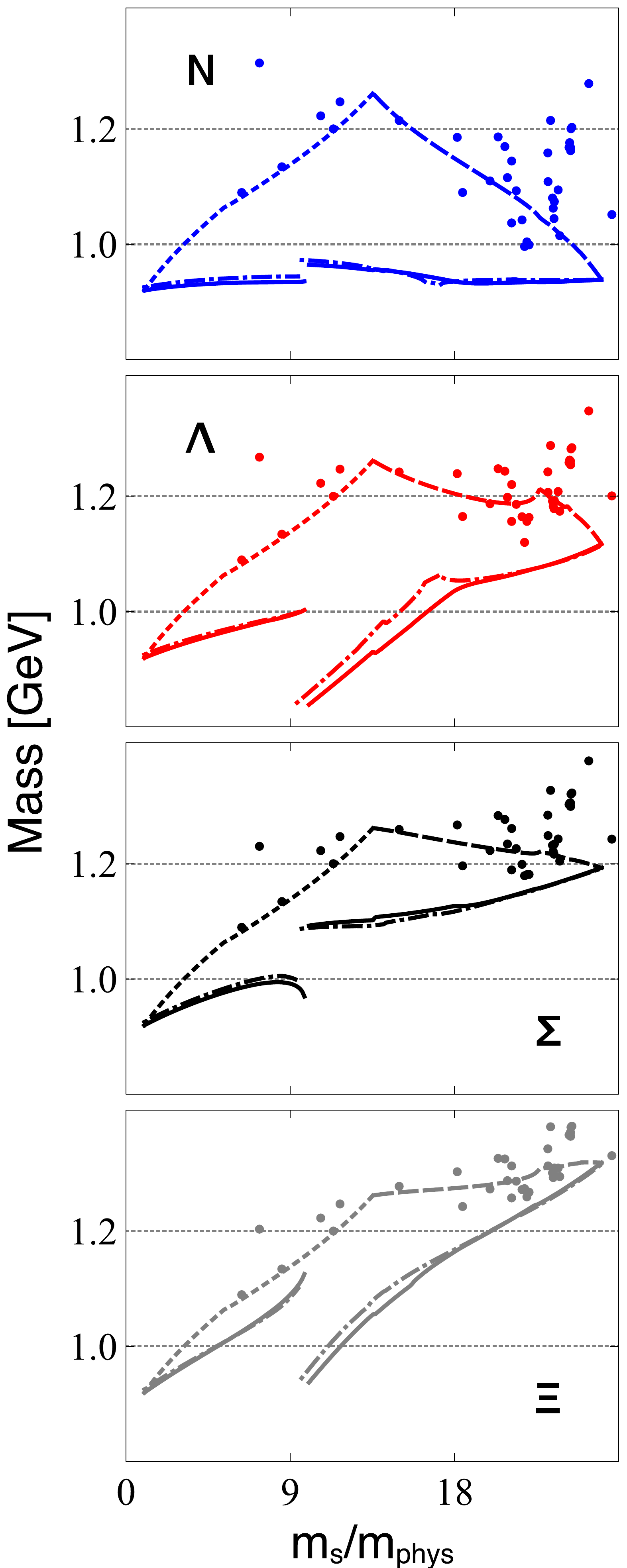}
\caption{Baryon octet masses as a function of the strange quark mass as explained in the text. }
\end{figure}

Results of two distinct global fits are collected in Tab. \ref{tab:Q1}. 
The error estimates for the low-energy parameters are based on statistical uncertainties at the one-sigma level. A complete access to the systematical error is beyond the scope of current analyses. Nevertheless,  we are confident that we derived a faithful representation 
of QCD in terms of its effective chiral Lagrangian (\ref{def-L1}, \ref{def-L2}). 
Both fits reproduce the set of baryon octet and decuplet masses on almost all QCD lattice ensembles with an accuracy of about 10 MeV. For technical details of our fit strategy we refer to our previous work \cite{Lutz:2018cqo}. 
Our Fit 1 and Fit 2 of Tab. \ref{tab:Q1} assume a residual theoretical uncertainty of 10 MeV and 5 MeV for each of the baryon masses  respectively. The total $\chi^2$ per degree of freedom is 1.06 and 1.70 in the two cases.  

We note that in our previous work \cite{Lutz:2018cqo} slightly incorrect values for the kaon masses on the ETMC ensembles \cite{Alexandrou:2013joa} were used. Published values are available 
only for the unitary case where the sea quark and valence quark masses are identical \cite{Lutz:2014oxa}. By  courtesy of
Constantia Alexandrou we can now consider the kaon masses as they are implied by the fine-tuned strange quark masses off their unitarity limit. Such strange quarks were used for the baryon masses in \cite{Alexandrou:2013joa}.

We carefully scrutinized the consistency of the various lattice data sets
and do find a quite convincing pattern \cite{Lutz:2018cqo}. Our current work is largely based on that
comprehensive study, with that only minor update of the ETMC data set \cite{Alexandrou:2013joa}. That update made the lattice data sets even more compatible. 
The only exception beeing the data on the  LHPC ensembles, which appear to set some tension. Like in our previous study \cite{Lutz:2018cqo} such data 
are reproduced with a somewhat larger uncertainty of 20 MeV only. 
Therefore, in our chisquare function such data points are assigned a systematic uncertainty of 20 MeV. We checked that our results are stable with respect to 
a complete omission of such data points.

As pointed out already in our previous works \cite{Lutz:2018cqo,Guo:2018kno} the low-energy constants, $2\,L_6 -L_4, 2\,L_8 - L_5$ and $L_8 + 3\,L_7$ 
can be determined quite accurately from a global fit to the baryon masses as measured on various QCD lattice ensembles. This is so since the baryon masses depend rather sensitively not only on the quark masses but also on the meson masses. 

In Fig. 1 we confront our results with the lattice data set for the baryon octet masses along the trajectory $m/m_{\rm phys}$. Finite volume effects are taken into account. 
Here we use the isospin averaged quark mass $m = m_u= m_d$. 
While the lattice data are shown by filled colored symbols, the chiral extrapolation result by corresponding open symbols displayed on top of the lattice points. Fit 1 results are systematically shown on top of Fit 2 results. 
The fact that almost always an open symbol covers its corresponding colored symbol confirms that the QCD lattice data are very well reproduced by our two fits. It is important to note that the 
lattice data set provides the baryon masses at various values of the strange quark mass. This has  decisive impact on the determination of the set of low-energy parameters.

Fig. 2  shows the baryon masses in the infinite volume limit as a function of $m_s/m_{\rm phys}$ based on three distinct scenarios. The first two are implied by $m= m_u=m_d=m_{\rm phys}$ (solid lines) or $m_u=m_d= m_s$ (dotted lines). The last one is implied by $ m + m_s =  (m+m_s)_{ \rm phys}$ (dashed lines). 
The differences of Fit 1 and Fit 2 are largest along
the $m= m_{\rm phys}$ trajectory, for which we show  the results of Fit 2 in terms of dot-dashed line. All other lines are with respect to Fit 1. 
For all trajectories we find a smooth $m_s$ dependence of the meson masses, but not of the baryon masses. On the $m= m_{\rm phys}$ trajectory we find a sizeable first order transition at $m_s/m_{\rm phys} \simeq $ 9.7 in the baryon masses. Once the strange quark mass exceeds its critical value stable baryonic matter is composed from strange particles, the $\Lambda$, rather than nucleons. In the region 
\begin{eqnarray}
 9.7 < m_s/{m_{\rm phys}}  < 14.8 \qquad \leftrightarrow \qquad  M_\Lambda < M_N \,,
 \label{def-gap}
\end{eqnarray}
normal baryonic matter does not exist in our preferred scenario. 
The figures are supplemented with a compilation of baryon masses as they are implied by our Fit 1 of the lattice data set, where now the masses are given in the finite box as set up by the lattice groups. Such points illustrate that so far lattice data provide very few direct constraints on hadron masses close to the $m = m_{\rm phys}$ trajectory.
We emphasize that all theory points shown in Fig. 2 correspond to the open-symbol points of Fig. 1 and therefore are as close to their 
lattice points as shown already in Fig. 1. The theory points are just displayed in a different manner.

A first order transition is implied also in the standard model scenario where the ratio $2\,m_s/(m_u+m_d)  \simeq 26$ takes its physical value. Along such a trajectory the light quark masses $m_u$ and $m_d$ are smaller than their physical values. Our findings are stable against one-sigma variations in the LEC.

\begin{table}[h]
\setlength{\tabcolsep}{2.5mm}
\renewcommand{\arraystretch}{1.5}
\begin{center}
\begin{tabular}{l|rrr}
                                       &  Fit 1      &  Fit 2       \\  \hline

$10^3\,( 2\,L_6 -L_4 )\, $             &  0.0402(${}_{-02}^{+12}$)    &  0.0401(${}_{-01}^{+24}$)    \\
$10^3\,( 2\,L_8 -L_5)\, $              &  0.1243(${}_{-07}^{+14}$)    &  0.1049(${}_{-43}^{+43}$)    \\
$10^3\,(L_8 + 3\,L_7)\, $              & -0.4866(${}_{-6}^{+2}$)      & -0.4818(${}_{-13}^{+09}$)    \\
$m_s/ m $                              &   25.91(${}_{-1}^{+0}$)      &   26.02(${}_{-2}^{+2}$)      \\ \hline
$M $[GeV]                              &  0.8662(${}_{-5}^{+6}$)      &  0.8726(${}_{-9}^{+4}$)      \\
$b_0 $[GeV$^{-1}$]                     & -0.7229(${}_{-71}^{+43}$)    & -0.7136(${}_{-65}^{+38}$)    \\
$b_D $[GeV$^{-1}$]                     &  0.0801(${}_{-09}^{+25}$)    &  0.0826(${}_{-7}^{+6}$)      \\
$b_F $[GeV$^{-1}$]                     & -0.3399(${}_{-35}^{+19}$)    & -0.3189(${}_{-30}^{+27}$)    \\
$g_0^{(S)} $[GeV$^{-1}$]               & -8.8678(${}_{-1664}^{+1446}$)& -8.5912(${}_{-1630}^{+1125}$)\\
$g_1^{(S)} $[GeV$^{-1}$]               &  0.8058(${}_{-385}^{+506}$)  &  0.6891(${}_{-785}^{+622}$)  \\
$g_D^{(S)} $[GeV$^{-1}$]               & -1.4485(${}_{-780}^{+522}$)  & -1.1512(${}_{-1252}^{+1413}$)\\
$g_F^{(S)} $[GeV$^{-1}$]               & -5.1101(${}_{-2227}^{+0985}$)& -4.7973(${}_{-1490}^{+1524}$)\\
$g_0^{(V)} $[GeV$^{-2}$]               & -0.3710(${}_{-2526}^{+1927}$)& -0.0446(${}_{-4507}^{+3671}$)\\
$g_1^{(V)} $[GeV$^{-2}$]               & -7.2709(${}_{-2337}^{+1724}$)& -6.6002(${}_{-2744}^{+3572}$)\\
$g_D^{(V)} $[GeV$^{-2}$]               &  10.002(${}_{-231}^{+369}$)  &  8.6215(${}_{-6432}^{+5680}$)\\
$g_F^{(V)} $[GeV$^{-2}$]               & -2.8688(${}_{-1715}^{+0984}$)& -2.6095(${}_{-2858}^{+2692}$)\\

\end{tabular}
\caption{Results for LEC in (\ref{def-L1}) based on two fit scenarios.
The low-energy constants $L_n$ are at the renormalization scale
$\mu = 0.77$ GeV. We use $f \simeq  92.4$ MeV and $F \simeq 0.48, D \simeq 0.75$ throughout this work.}
\label{tab:Q1}
\end{center}
\end{table}

\section{Summary and conclusions}

We have shown that QCD lattice data on the baryon masses are compatible  with a first order transition along the trajectory where the strange quark mass goes from its chiral limit to its physical value. Our global fits predict an anomalous sector of QCD where stable baryonic matter would be composed of $\Lambda$ or $\bar \Lambda$ particles rather than nucleons or anti-nucleons. Such dark matter may exist if the Higgs potential in extensions of the Standard Model supported two almost degenerate minima, with the exotic one implying a strange quark close to its critical first order transition value \cite{Lutz:2019gmv}. 

We suggest further studies to consolidate our results. In particular it would be useful to construct an effective potential approach that leads to our set of gap equations for the baryon octet and decuplet masses.

\begin{acknowledgements}
John Bulava and Madeleine Soyeur are acknowledged for stimulating discussions.
Y. H. received partial support from Suranaree University of Technology, Office of the Higher Education Commission under NRU project of Thailand (SUT-COE: High Energy Physics and Astrophysics) and SUT-CHE-NRU (Grant No. FtR.11/2561).
\end{acknowledgements}


\bibliographystyle{spphys}
\bibliography{literature}

\end{document}